\documentclass[useAMS,usenatbib,usegraphicx]{mn2e}

\title{A phenomenological model of galaxy clusters}
\author[Y.~Ascasibar and J.~M.~Diego]
{
  Y.~Ascasibar$^1$\thanks{E-mail: yago@aip.de} and J.~M.~Diego$^2$\\
  $^1$ Astrophysikalisches Institut Potsdam, An der Sternwarte 16, Potsdam 14482, Germany\\
  $^2$ Instituto de F\'\i sica de Cantabria, Avenida de los Castros s/n, Santander 39005, Spain
}

\newcommand{\be}{\begin{equation}}
\newcommand{\ee}{\end{equation}}

\newcommand{\dd}{{\rm d}}

\newcommand{\dm}{_{\rm dm}}
\newcommand{\gas}{_{\rm gas}}
\newcommand{\cc}{_{\rm c}}
\newcommand{\rt}{r_{\rm t}}
\newcommand{\ac}{a\cc}

\begin{document}

\maketitle

\begin{abstract}
We present a simple model to describe the dark matter density, the gas density, and the gas temperature profiles of galaxy clusters.
Analytical expressions for these quantities are given in terms of only five free 
parameters with a clear physical meaning: the mass $M$ of the dark matter halo 
(or the characteristic temperature $T_0$), the characteristic scale radius $a$, 
the cooling radius in units of $a$ ($0<\alpha<1$), the central temperature in units of 
$T_0$ ($0<t<1$), and the asymptotic baryon fraction in units of the cosmic value 
($f\sim1$).
It is shown that our model is able to reproduce the three-dimensional density and 
temperature profiles inferred from X-ray observations of real clusters within a 
20 per cent accuracy over most of the radial range.
Some possible applications are briefly discussed.
\end{abstract}


  \section{Introduction}
  \label{secIntro}

Observations of galaxy clusters provide invaluable information about the universe and its evolution.
The abundance of galaxy clusters as a function of temperature and redshift, their baryon fraction, 
the X-ray luminosity, the Sunyaev-Zeldovich and lensing effects have often been used as cosmological 
probes \citep[see e.g.][for a recent review]{Voit05}.
In particular, they constrain the matter density, the slope and normalization of the power spectrum of primordial fluctuations, and the equation of state of dark energy.
Thanks to the advent of high-resolution X-ray observatories, we can now study the structure of the intracluster medium (ICM) with unprecedented accuracy.
For the brightest objects, the gas density and temperature profiles can be directly inferred from the observed data, assuming approximate spherical symmetry.
However, it is not uncommon that these observations have a limited field of view, and the profiles have to be fitted by some analytical model in order to extrapolate them outwards.
In fainter systems, the use of analytical models makes possible a more accurate determination of the cluster parameters.
In the most extreme case, the models actually play a crucial role in the very detection of the faintest cluster candidates.
It is therefore important to have a simple analytical description of the ICM that captures the most relevant features of the density and temperature profiles of the X-ray emitting gas. 

Historically, one of the most popular options is the so-called $\beta$-model \citep{CavaliereFusco76}, where the gas density is given in terms of three free parameters (the normalization $\rho_0$, a core radius $r\cc$, and the exponent $\beta$),
\be
  \rho\gas(r)=\frac{\rho_0}{ \left[1+(r/r\cc)^2\right]^{3\beta/2} }
  \label{eqBM}
\ee
The $\beta$-model has been widely used over the years to describe the radial structure of the ICM.
However, it is well known that equation~(\ref{eqBM}) fails to provide a consistent fit over the whole radial range.
Moreover, the gas is assumed to be isothermal, which is certainly not consistent with the observed temperature profiles.
Finally, for a gas in hydrostatic equilibrium, the dark matter distribution underlying a $\beta$-model would be
\be
  \rho\dm(r)\propto\frac{1+(r/r\cc)^2/3}{ \left[1+(r/r\cc)^2\right]^2 }
\ee
which tends to a constant value at the centre.
Such a `cored' density profile is in strong disagreement with the results of numerical simulations, as well as with most recent observational estimates of the mass distribution in galaxy clusters. 

Polytropic models, where the density and temperature of the gas are related by the effective equation of state
\be
  \frac{\rho\gas(r)}{\rho_0} = 
  \left[ \frac{T(r)}{T_0} \right]^n
\ee
have been shown to provide a much better description of the ICM for $n\sim5$.
Nevertheless, the temperature profile predicted by these models tends to increase steadily towards the cluster centre, whereas both numerical experiments and observations of real systems show that radiative cooling makes the temperature drop in the central regions of the ICM.
On the other hand, the central gas density is often observed to become a power-law rather than a constant density core.

This has motivated the introduction of more elaborate analytical models, including additional free parameters in order to describe a wider range of possible behaviours \citep[see e.g.][hereafter V06, and references therein]{Vikhlinin+06}.
In particular, these authors propose a model, characterized by 17 independent free parameters, that accurately reproduces currently available X-ray data.
According to the V06 model, the three-dimensional gas density profile would be described by
\begin{eqnarray}
\rho\gas^2(r) & = &
  \nonumber
  \frac{ \rho_1^2\ (r/r_{\rm c1})^{-\alpha} }
       {
	\left[ 1+(r/r_{\rm c1})^2 \right]^{3\beta_1-\alpha/2}
	\left[ 1+(r/r_{\rm s})^3 \right]^{\epsilon/3} }\\
  & &
  + \frac{ \rho_2^2 }{ \left[ 1+(r/r_{\rm c2})^2 \right]^{3\beta_2} }
 \label{eqRhoV06}
\end{eqnarray}
while the gas temperature is modelled as
\be
T(r) =
 \frac{T_{\rm min} + T_0(r/r\cc)^{a\cc}}{1+(r/r\cc)^{a\cc}}
 \frac{(r/\rt)^{-a}}{\left[1+(r/\rt)^b\right]^{c/b}}
 \label{eqTV06}
\ee

Equations~(\ref{eqRhoV06}) and~(\ref{eqTV06}) have great freedom and can provide a good fit to the observed density and temperature profiles, both for the inner and outer regions of the ICM.
The main disadvantage, though, is that their 17 parameters are strongly correlated, and thus there are many degeneracies in their best-fitting values.
This would be a relatively minor problem (e.g. computational cost) if our only goal was to reproduce the observational data, but it becomes of critical importance when one attempts to extrapolate outside the observed region or when the number of photons from the object under investigation is too low to obtain reliable profiles.
In these cases, robustness becomes more important than flexibility, and a model with fewer parameters is preferable in order not to over-fit the available data.
A simple, robust model can be extremely helpful, for instance, in cosmological studies, where the observed number counts of galaxy clusters as a function of temperature or luminosity need to be connected with the underlying mass function.
Such a model would also be of great interest for multi-wavelength analysis, where data with very different errors are combined in order to recover the three-dimensional structure of the object.

Here we present an analytical model that is able to reproduce the complex behaviour of the gas density and temperature profiles observed in real galaxy clusters by using only five free parameters, all of which have a clear physical meaning.
The model and its parameters are described in detail in Section~\ref{secModel}.
We compare it with observational data in Section~\ref{secResults}, where it is shown that our simple model fits all the observable X-ray properties of the ICM, while being consistent with our current knowledge of the structure of dark matter haloes.
In Section~\ref{secApplic}, we consider three possible applications, namely the set up of initial conditions in numerical experiments, the construction of optimal filters for X-ray detection, and the combined analysis of X-ray and Sunyaev-Zel'dovich data.
Conclusions are briefly summarized in Section~\ref{secConclus}.

  \section{Model description}
  \label{secModel}


\subsection{Dark matter}

Perhaps one of the best known results of cosmological N-body simulations is that the radial density profiles of dark matter haloes can be well fitted by a relatively simple analytical function with very few parameters \citep[e.g.][]{NFW97}.
The precise shape of such function, particularly near the centre, is still a matter of heated debate \citep[see e.g.][for a recent discussion]{Merritt+06}, but there is general agreement in that it should be shallower than isothermal ($\rho\propto r^{-2}$) as $r\to 0$ while significantly steeper as $r\to\infty$.

In order to model the cluster's dark matter halo, we use a \citet{Hernquist90} density profile,
\be
\rho(r)=\frac{M}{2\pi a^3}\frac{1}{r/a(1+r/a)^3}
\label{eqRhoH}
\ee
where $M$ denotes the total mass and $a$ is a characteristic scale length.
The cumulative mass inside radius $r$ is given by
\be
M(r)=M\left(\frac{r/a}{1+r/a}\right)^2,
\label{eqMH}
\ee
and the gravitational potential is simply
\be
\phi(r)=\frac{GM}{a+r},
\ee
where $G$ is Newton's constant.
Analytical expressions for other quantities, such as the velocity dispersion, distribution function or projected surface density, can be found in \citet{Hernquist90}.

For the sake of simplicity, we will assume that equations~(\ref{eqRhoH}) and (\ref{eqMH}) correspond to the \emph{total} density and mass, respectively, i.e. the sum of the dark and baryonic components.
The dark matter profiles can be trivially obtained by subtracting the contribution of the gas.


\subsection{Polytropic equation of state}

Non-radiative gasdynamical simulations show that, in the absence of additional physics, the ICM of relaxed clusters can be approximately described as a polytropic gas in hydrostatic equilibrium with the gravitational potential created by the dark matter \citep[e.g.][]{Ascasibar+03}.
Under these conditions, one can derive analytical expressions for both the gas temperature
\be
T(r)=\frac{T_0}{1+r/a}
\label{eqTpolyt}
\ee
and density
\be
\rho_g(r)=\frac{\rho_0}{(1+r/a)^n}
\ee
profiles, where $T_0$ and $\rho_0$ correspond to the central values, and $n$ is the effective polytropic index.
Hydrostatic equilibrium also imposes the mass-temperature relation
\be
(n+1)\frac{kT_0}{\mu m_{\rm p}}=\frac{GM}{a}
\label{eqMT}
\ee
where $k$ is the Boltzmann constant, $m_{\rm p}$ denotes the proton mass, and $\mu\simeq0.6$ is the molecular weight of the gas.
In order to obtain a constant baryon fraction at large radii, we set $n=4$.

\begin{figure*}
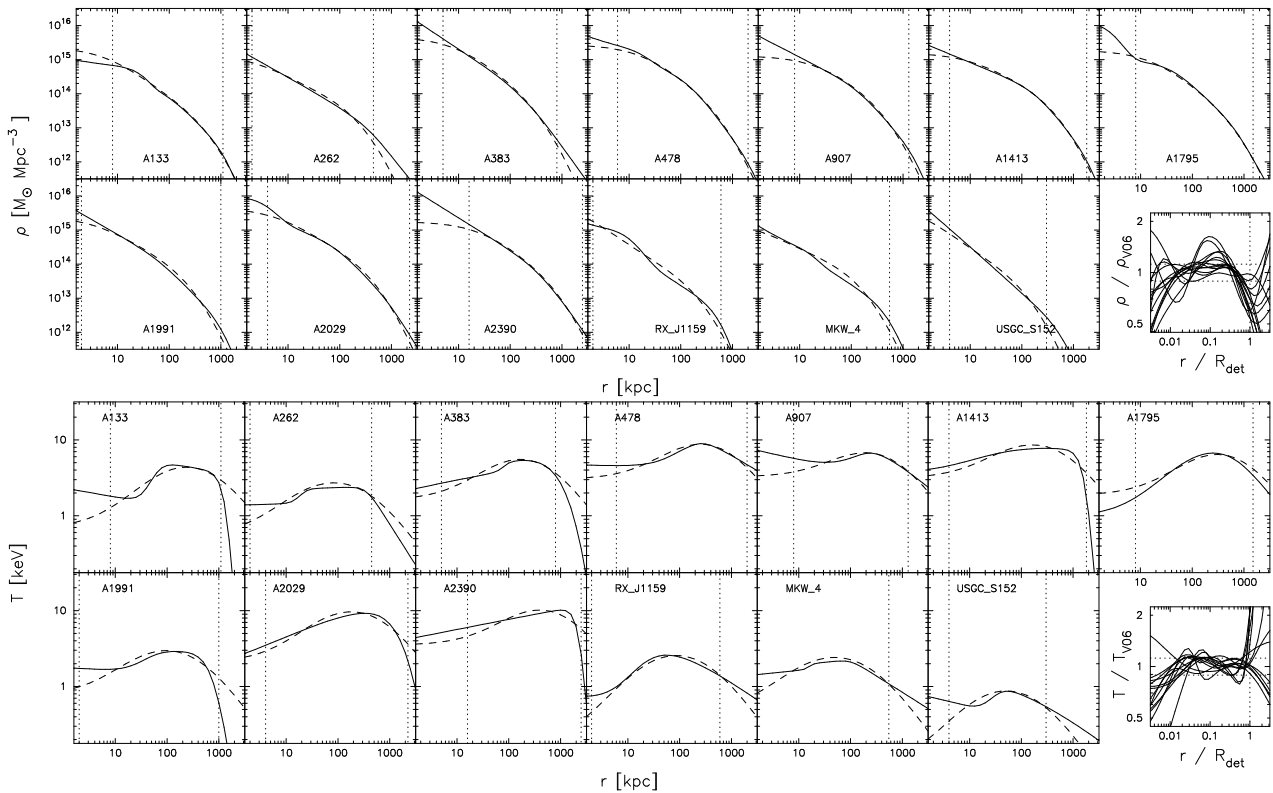

\includegraphics[width=.95\textwidth]{Rho.eps}\\
\includegraphics[width=.95\textwidth]{T.eps}
\caption
{
Three-dimensional density (top) and temperature (bottom) profiles for the 13 objects in V06 (solid lines), compared to the best fit obtained with our model (dashed lines).
Vertical dotted lines indicate the radial range used for the fit (the same as in V06), and values of the best-fitting parameters are given in Table~\ref{tabFit}.
Last panels show the fractional deviation between our model and V06 (horizontal dotted lines indicate $\pm0.05$ in logarithmic scale, and a vertical dotted line is drawn at the \emph{Chandra} detection radius $R_{\rm det}$).
}
\label{figFit}
\end{figure*}


\subsection{Cool core}

The polytropic model roughly agrees with observational results \citep[e.g.][]{Markevitch98}, but the observed temperature profiles often feature a central drop that is not well reproduced by equation~(\ref{eqTpolyt}).
Therefore, we introduce a modified temperature
\be
T(r)=\frac{T_0}{1+r/a}~\frac{t+r/\ac}{1+r/\ac},
\label{eqT}
\ee
where $0<t<1$ is a free parameter that measures the amount of central cooling with respect to the polytropic solution, and $a\cc$ reflects the cooling radius below which the effect is important.
Physically, one expects $a\cc<a$, so we consider the parametrization $\alpha=a\cc/a$ with $0<\alpha<1$.

Dropping the polytropic assumption, we can substitute this expression into the hydrostatic equilibrium equation and compute the corresponding gas density,
\be
\frac{\rho_{\rm gas}(r)}{\rho_0} =
  \left( \frac{1+r/a}{t\alpha+r/a} \right)^{1+\frac{\alpha-t\alpha}{1-t\alpha}(n+1)}
  \frac{\alpha+r/a}{\left(1+r/a\right)^{n+1}}
\label{eqRho}
\ee
where the normalization $\rho_0$ can be expressed in terms of the cosmic baryon fraction,
\be
\rho_0 = f \frac{\Omega_{\rm b}}{\Omega\dm} \frac{M}{2\pi a^3}
\label{eqFb}
\ee
with $f\sim1$.

  \section{Comparison with real clusters}
  \label{secResults}

In order to test whether our simple prescription provides an adequate model of galaxy clusters, we consider the sample of 13 low-redshift, relaxed objects studied by V06.
Rather than fitting the raw observational data, we simply attempt to reproduce the three-dimensional density and temperature profiles, described by equations~(\ref{eqRhoV06}) and (\ref{eqTV06}) with values of the $17$ free parameters according to tables~2 and 3 of V06.
We account for the errors in the deprojected quantities by assuming a log-normal distribution with $\sigma_\rho=\sigma_{\rm T}=0.05$.

For each object, we fit the three-dimensional profiles within the radii $0.2R_{\rm min}$ and $R_{\rm det}$, defined in V06.
We vary $T_0$, $t$, $a$, $\alpha$, and $f$ in 30 logarithmic steps and compute the reduced chi-square as
\be
 \chi^2 = \frac{ \chi^2_\rho + \chi^2_{\rm T} }{2N-5}
\ee
with
\be
\chi^2_\rho =
  \frac{1}{\sigma_\rho^2}
  \sum_{i=1}^{N}
  {\left[
    \log\left(\frac{\rho\gas(r_i)}{\rho_{\rm V06}(r_i)}\right)
  \right]^2}
\ee
and
\be
\chi^2_{\rm T} =
  \frac{1}{\sigma_{\rm T}^2}
  \sum_{i=1}^{N}
  {\left[
    \log\left(\frac{T(r_i)}{T_{\rm V06}(r_i)}\right)
  \right]^2}
\ee
The gas density and temperature are evaluated at $N=30$ points where the radius $r_i$ is also increased logarithmically.
We assume a cosmic baryon fraction $\Omega_{\rm b}/\Omega\dm=0.133$ and we relate the electron and proton number densities to the gas density as $n_{\rm e}= n_{\rm p} \approx \rho\gas/m_{\rm p}$.

\begin{table}
\begin{tabular}{rrrrrr}
\hline
$T_0$ (keV) & $t$~~ & $a$ (kpc) & $\alpha$~~ & $f$~~ & $\chi^2$~~ \\
\hline
  7.2487 & 0.0949 &  840.9622 &  0.0949 &  0.7249 & 1.1217 \\
  4.0013 & 0.1372 &  400.1251 &  0.0547 &  0.6248 & 2.4631 \\
  9.7565 & 0.1649 &  538.5525 &  0.1372 &  0.6248 & 1.3281 \\
 15.2349 & 0.1982 &  840.9622 &  0.1372 &  0.8410 & 0.8646 \\
 11.3190 & 0.2864 &  724.8703 &  0.1372 &  0.8410 & 1.5131 \\
 13.1318 & 0.2382 &  724.8703 &  0.0790 &  1.1319 & 0.9807 \\
 11.3190 & 0.1649 &  975.6467 &  0.1372 &  0.7249 & 0.8148 \\
  4.6421 & 0.1649 &  400.1251 &  0.0790 &  0.8410 & 3.9217 \\
 15.2349 & 0.1372 &  724.8703 &  0.0790 &  1.1319 & 1.6926 \\
 17.6749 & 0.1982 & 1313.1818 &  0.1372 &  1.1319 & 1.2798 \\
  4.0013 & 0.0378 &  344.8893 &  0.0657 &  0.3449 & 4.8305 \\
  3.4489 & 0.1372 &  256.2403 &  0.0455 &  0.4642 & 6.5543 \\
  1.4144 & 0.0547 &  190.3773 &  0.0790 &  0.3449 & 3.3169 \\
\hline
\end{tabular}
\caption
{
Best-fitting values and reduced $\chi^2$ for each cluster.
}
\label{tabFit}
\end{table}

Values of the best-fitting parameters are given in Table~\ref{tabFit}, and the resulting gas density and temperature profiles are plotted for each object in Figure~\ref{figFit}, where the last panels show the fractional deviations with respect to the V06 model. For all systems, the discrepancy is of the order of $10-20$ per cent within the fitted range, consistent with our adopted estimate of the error bars, $\sigma_\rho=\sigma_{\rm T}=0.05$ (see also the reduced $\chi^2$ values in Table~\ref{tabFit}).
Outside this range, the model tends to yield densities and temperatures at small radii that are systematically lower than those measured by V06.
Although this might actually be correct for some systems (see the observational data points in V06), it will not be so in others.
It is likely that structure on small scales (e.g. cold and shock fronts, the very presence of a central galaxy) breaks down the assumptions of perfect spherical symmetry and hydrostatic equilibrium, making any simple model inadequate to describe the ICM (and maybe even the dark matter potential) in the innermost part of the cluster, and indeed the objects that are worst described by our model display one or more inflection points in their profiles.
At large radii, the density profile given by equation~(\ref{eqRho}) seems to be steeper than the results of V06.
In order to test the ability of our model to infer the cluster properties from data of poorer quality, restricted to a smaller field of view, we repeated our analysis considering only $N=3$ radii between $R_{\rm min}$ and $0.5R_{\rm det}$.
The accuracy of the recovered profiles was similar within the fitted region, but the extrapolation towards large radii was not entirely reliable, with errors of the order of a factor of two or above in the worst cases.

\begin{figure}
\includegraphics[width=8cm]{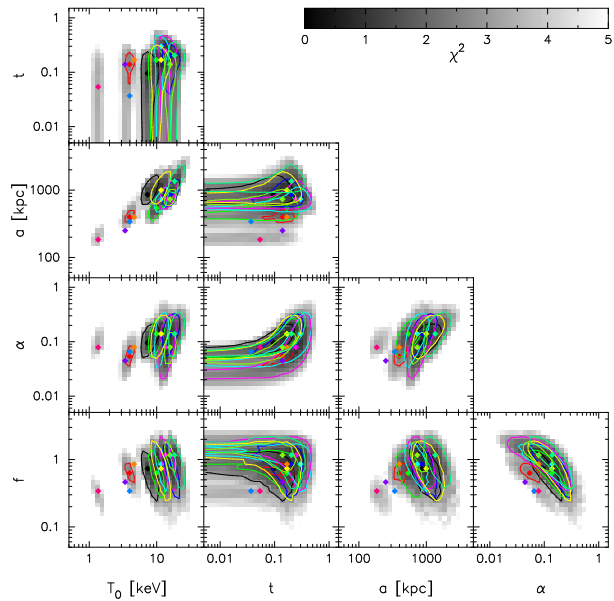}
\caption
{
Maps of minimum $\chi^2$ for all clusters.
Dots mark the best fit for each object, and contours are drawn at $\chi^2=3$.
}
\label{figLikeli}
\end{figure}

The shape of the likelihood function, or, more precisely, $\chi^2(T_o,t,a,\alpha,f)$, is studied in Figure~\ref{figLikeli}. For each cluster, a diamond marks the best-fitting parameters and a contour is drawn at $\chi^2=3$. The underlying grayscale maps show the projected $\chi^2$, marginalized over all clusters and all possible values of the other parameters (i.e. the minimum $\chi^2$ attained at each point by any object).
Our results suggest that the likelihood function is well behaved, in the sense that it displays well-defined, unique maxima for each cluster, but it seems to be significantly skewed (i.e. the best fit does not always coincide with the geometrical centre of the contour) given our particular choice of variables.

For each object, there is only a relatively mild degeneracy between the best-fitting values of the five free parameters of our model, although a certain mutual dependence is obvious in several pairs (for instance, those involving $\alpha$, $t$ and $f$).
Considering the whole sample, we find strong correlations (i.e. scaling relations) between most parameters and the mass of the cluster, or equivalently $T_0$.
Although more observational data and/or numerical simulations would be required in order to make a quantitative statistical assessment, it is interesting to note that these correlations could be exploited to reduce the number of free parameters in the model even further.
If all parameters could indeed be expressed as a function of $T_0$, our model would effectively have one single free parameter, specifying the overall scale of the cluster.

  \section{Applications}
  \label{secApplic}

Having a simple analytical model to describe the three-dimensional distribution of gas, temperature and matter in galaxy clusters can be useful in many respects.
On the one hand, it may serve as a base to make theoretical predictions (e.g. to study the effect of a given perturbation on the ICM of a relaxed cluster).
On the other hand, it also helps to interpret observational data (e.g. to estimate the mass of the cluster, or the temperature profile, with a small number of photons).
In this section we briefly discuss three examples of possible applications of a model like the one presented in this paper.

\subsection{Numerical simulations}

Our model can be used, for instance, to set up the initial conditions in idealized numerical experiments \citep[and, in fact, it has already been used for this purpose in][]{AscasibarMarkevitch06}.
In a few words, the procedure to generate a synthetic cluster is as follows:
\footnote{Computer code is available upon request.}

First, the radius of each particle is obtained by generating a uniform random number $\mu$ between 0 and 1 and inverting the appropriate (gas or dark matter) mass profile, i.e. solving for $M(r)=\mu M_{\rm x}$, where $M_{\rm x}$ denotes the corresponding total mass.
Angular coordinates $\phi$ and $\cos\theta$ are uniform random numbers in the range $[0,2\pi]$ and $[-1,1]$, respectively.

Gas temperature and density are given by equations~(\ref{eqT}) and~(\ref{eqRho}) as a function of radius.
When modelling a composite system containing several objects, the density of each particle is computed as $\rho=\sum\rho(r_i)$, where $r_i$ is the distance to object $i$, and temperatures are set according to $T=\sum\rho(r_i)T(r_i)/\rho$.
Velocities are given by ${\bf v}=\sum\rho(r_i){\bf v_i}/\rho$, with ${\bf v_i}$ being the centre of mass velocity of the $i$-th object.

For collisionless dark matter particles, each object is completely independent from the others. Velocities with respect to the relevant centre of mass are assigned from the probability distribution
\be
p(v)\,\dd v= \frac{4\pi}{\rho(r)}\ f( {v^2}/{2}+\Phi(r) )\ v^2 \dd v
\label{eqDMvel}
\ee
where $\Phi(r)$ denotes the gravitational potential, and the distribution function $f(E)$ is computed by using Eddington's formula \citep{BT},
\be
f(E) = 2^{-3/2}\pi^{-2} \frac{\dd}{\dd E}
\left[\ \int_E^0 (\Phi-E)^{-1/2}\frac{\dd\rho}{\dd\Phi}\ \dd\Phi\ \right]
\ee
The probability distribution~(\ref{eqDMvel}) is sampled by means of the von Neumann rejection algorithm \citep[see e.g.][]{NR}.
We generate a tentative velocity $v$, uniformly distributed between 0 and $v_{\rm max}=\sqrt{-2\Phi(r)}$, and an auxiliary random number $p$ between 0 and $v_{\rm max}^2f(v_{\rm max})$.
The velocity is accepted only if $p<v^2f(v)$.
Otherwise, two new random numbers are generated for $v$ and $p$ until a value is finally accepted for the velocity of the particle.
As for positions, the angular coordinates of the velocity are obtained from two uniform random variables, $0<\phi<2\pi$ and $-1<\cos\theta<1$.

\subsection{Optimal filtering}

A major issue in the construction of galaxy cluster samples is the optimization of the detection algorithm.
One of the most successful approaches is based on the use of wavelets \citep[see e.g.][]{Rosati+95,Lazzati+99}.
Simple bases, like the Mexican Hat wavelet, are optimal when the underlying signal is Gaussian and the background has a $k^{-2}$ power spectrum, but neither condition is met by cluster X-ray data.
The underlying signal would be better described by our model, and the background can be modelled as a random Poisson variable whose normalization varies from pointing to pointing. In the case of the Sunyaev-Zel'dovich effect, some authors have improved the detection algorithms by adopting optimal filters instead of wavelets and assuming that the gas is described by a $\beta$-model \citep[][]{Herranz+02}.
As mentioned in the introduction, the $\beta$-model does not capture the complex behaviour of the gas density and temperature profiles; moreover, it fails to provide a good description of the ICM in the outer regions, to which the Sunyaev-Zel'dovich effect is also sensitive.
Therefore, an optimal filter based on our simple model could help to improve the detection rate in both the X-ray and millimeter bands.

\begin{figure}
\includegraphics[width=.45\textwidth]{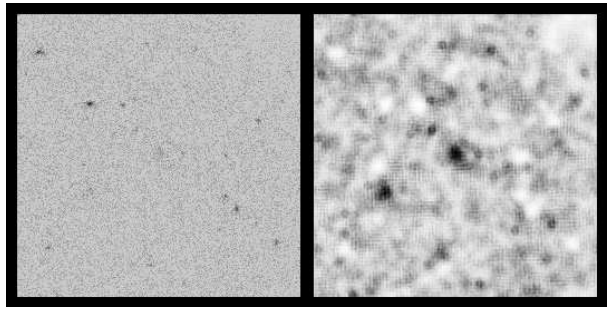}
\caption
{
Optimal filtering of X-ray data. The left panel shows a real XMM image 
(17 arcmin on a side) with a superposed simulated cluster at the center 
of the image. The right image shows the optimally filtered image after point source 
subtraction. The cluster at the center is now evident. Note also a cluster candidate 
to the southwest of the central object. 
}
\label{figXMM}
\end{figure}

For this purpose, simplicity is a major concern, since the shape and scale of the filter depend on the values of the free parameters.
Our model provides an accuracy comparable to that of V06, at the expense of only one extra parameter with respect to the $\beta$-model.
As discussed above, the correlations observed in Figure~\ref{figLikeli} suggest that
it might be possible to define an optimal filter in terms of only one single scale parameter.
Although exploring in detail the performance of such a filter is well beyond the scope of the present work, we show in Figure~\ref{figXMM} an example of its application to cluster detection in X-ray data.
A simulated cluster (based on one of the 13 models) has been added to real XMM data (see left panel of figure \ref{figXMM}). 
After subtracting the brightest sources, we build an optimal filter based on the model of the cluster and the XMM background. The result is shown on the right panel of figure \ref{figXMM}, where the simulated object can be seen in the center of the image, as well as another cluster candidate to its left.

\subsection{Multi-wavelength 3D deprojection}

\begin{figure}
\includegraphics[width=.45\textwidth]{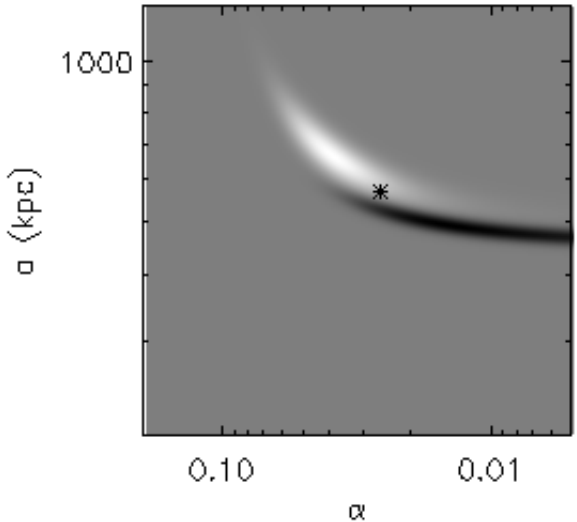}
\caption
{
Likelihoods in the $\alpha-a$ space when the model is compared to two simulated noisy data sets: an X-ray (black) and a Sunyaev-Zel'dovich (white) image.
The asterisk marks the position of the fiducial model used to make the 
simulated data. The other 3 parameters where not varied.
}
\label{figSZ_XR}
\end{figure}

There is already a number of galaxy clusters that have been observed in different wavebands, and this number is expected to increase dramatically during the forthcoming years.
Since different observations probe the gas or matter distribution in different ways, a combined analysis provides extremely tight constraints on the three-dimensional properties of the cluster.
For instance, the X-ray emissivity of the plasma, the magnitude of the thermal Sunyaev-Zel'dovich effect at millimeter bands and the convergence or shear of gravitational lensing (measured from optical data) are proportional to the integral along the line of sight of the gas density squared, the gas pressure, and the mass density, respectively.
Taking into account all these observables, it is possible to reconstruct the three-dimensional structure of clusters with an extraordinary accuracy.
Here we illustrate the problem with a simple example, involving two observations of the same object (an X-ray image and a Sunyaev-Zel'dovich image).
Both datasets were simulated by adding uniform random noise to one realization of our model, and likelihoods on the $\alpha-a$ plane were derived in each case.
For the sake of simplicity, the remaining three parameters have been fixed to the values used to generate the data.
As can be seen in Figure~\ref{figSZ_XR}, each data set is slightly biased with respect to the fiducial model (asterisk).
When both results are combined together, the best-fitting $\alpha$ and $a$ are much closer to their true values.

An interesting possibility, further along this direction, would be to recover the triaxial structure of the dark matter halo.
Although the model we described here is spherically symmetric, it can be trivially modified to account for a triaxial gas distribution.
The shape of the gravitational potential and the density of the dark matter halo can then be obtained from the hydrostatic equilibrium condition.
Since the procedure involves second derivatives, it is extremely sensitive to the details of the gas distribution.
It is important that the model captures these details accurately, while at the same time it must be simple enough to filter the high-frequency noise, which otherwise would dominate the final result.

  \section{Conclusions}
  \label{secConclus}

We have presented an analytical model of galaxy clusters based on spherical symmetry and hydrostatic equilibrium.
Our model is fully specified by equations~(\ref{eqRhoH}), (\ref{eqMT}), (\ref{eqT}), (\ref{eqRho}), and (\ref{eqFb}), and its five parameters have a well-defined physical interpretation: $M$ is the mass of the system, $a$ is a characteristic scale length, $0<\alpha<1$ is the cooling radius in units of $a$, $0<t<1$ is the central temperature in units of $T_0$, and $f\sim1$ is the asymptotic baryon fraction in units of the cosmic value.

Its main advantage with respect to previous work is the reduced number of free parameters.
Besides a more straightforward interpretation of the results, describing the observations in terms of a few parameters is also very convenient from a computational point of view, since (depending on the details of the algorithm) the complexity of the fitting procedure scales roughly exponentially with the dimensionality of the parameter space.
An additional advantage are the smaller degeneracies, which make both the inferred profiles and the best-fitting values of the parameters more robust.
This, in turn, makes possible to investigate correlations (scaling relations) that reduce the effective number of free parameters.
Our results suggest that, in fact, it might be possible that the whole structure of a galaxy cluster could be completely determined by just one single characteristic scale.
This very interesting possibility needs to be investigated further with more data. If confirmed, it would provide a crucial improvement to our current ability to model -- and understand -- the internal structure of galaxy clusters.

Comparing our model to the three-dimensional density and temperature profiles obtained from high-resolution X-ray observations, we find that it provides an accurate description of the intracluster medium.
In particular, the fractional deviations with respect to the more elaborate model proposed by V06 are comparable to the measurement errors (of the order of $10-20$ per cent).
The systems that feature high values of the reduced $\chi^2$ usually display a complex behaviour that may be associated to departures from perfect hydrostatic equilibrium, but a much more detailed study would be required in order to test this hypothesis.

We have discussed a few possible applications of the model presented here.
First, it provides a simple description of galaxy clusters to set up the initial conditions in idealized numerical experiments.
Second, it can be helpful in the detection and characterization of observed systems, specially when the signal-to-noise ratio is not extremely large.
Finally, a simple model of the matter and gas distribution makes possible to reconstruct the three-dimensional structure of the cluster from multiwavelength data.
The construction of an optimal filter for cluster detection based on our model, as well as the combination of X-ray and Sunyaev-Zel'dovich information, will be the subject of future work.


\section*{Acknowledgments}

This work has been funded by the Spanish \emph{Ministerio de \mbox{Educaci\'on} y Ciencia}, under project AYA2006-06266.
YA would like to thank A. Vikhlinin for useful discussions, and the \emph{Instituto de F\'\i sica de Cantabria} for their kind hospitality.
JMD benefits from a \emph{Ram\'on y Cajal} contract from the \emph{Ministerio de Educaci\'on y Ciencia}.

 \bibliographystyle{mn2e}
 \bibliography{Profile}

\begin{thebibliography}{}

\bibitem[\protect\citeauthoryear{{Ascasibar} \& {Markevitch}}{{Ascasibar} \&
  {Markevitch}}{2006}]{AscasibarMarkevitch06}
{Ascasibar} Y.,  {Markevitch} M.,  2006, \apj, 650, 102

\bibitem[\protect\citeauthoryear{{Ascasibar}, {Yepes}, {M{\" u}ller} \&
  {Gottl{\" o}ber}}{{Ascasibar} et~al.}{2003}]{Ascasibar+03}
{Ascasibar} Y.,  {Yepes} G.,  {M{\" u}ller} V.,    {Gottl{\" o}ber} S.,  2003,
  \mnras, 346, 731

\bibitem[\protect\citeauthoryear{{Binney} \& {Tremaine}}{{Binney} \&
  {Tremaine}}{1987}]{BT}
{Binney} J.,  {Tremaine} S.,  1987, {Galactic dynamics}.
Princeton, NJ, Princeton University Press, 1987, 747 p.

\bibitem[\protect\citeauthoryear{{Cavaliere} \& {Fusco-Femiano}}{{Cavaliere} \&
  {Fusco-Femiano}}{1976}]{CavaliereFusco76}
{Cavaliere} A.,  {Fusco-Femiano} R.,  1976, \aap, 49, 137

\bibitem[\protect\citeauthoryear{{Hernquist}}{{Hernquist}}{1990}]{Hernquist90}
{Hernquist} L.,  1990, \apj, 356, 359

\bibitem[\protect\citeauthoryear{{Herranz}, Sanz, Hobson, Barreiro, Diego,
  {Mart\'\i nez-Gonz\'alez} \& {Lasenby}}{{Herranz} et~al.}{2002}]{Herranz+02}
{Herranz} D.,  Sanz J.~L.,  Hobson M.~P.,  Barreiro R.~B.,  Diego J.~M.,
  {Mart\'\i nez-Gonz\'alez} E.,    {Lasenby} A.~N.,  2002, \mnras, 336, 1057

\bibitem[\protect\citeauthoryear{{Lazzati}, {Campana}, {Rosati}, {Panzera} \&
  {Tagliaferri}}{{Lazzati} et~al.}{1999}]{Lazzati+99}
{Lazzati} D.,  {Campana} S.,  {Rosati} P.,  {Panzera} M.~R.,    {Tagliaferri}
  G.,  1999, \apj, 524, 414

\bibitem[\protect\citeauthoryear{{Markevitch}, {Forman}, {Sarazin} \&
  {Vikhlinin}}{{Markevitch} et~al.}{1998}]{Markevitch98}
{Markevitch} M.,  {Forman} W.~R.,  {Sarazin} C.~L.,    {Vikhlinin} A.,  1998,
  \apj, 503, 77

\bibitem[\protect\citeauthoryear{{Merritt}, {Graham}, {Moore}, {Diemand} \&
  {Terzi{\'c}}}{{Merritt} et~al.}{2006}]{Merritt+06}
{Merritt} D.,  {Graham} A.~W.,  {Moore} B.,  {Diemand} J.,    {Terzi{\'c}} B.,
  2006, \aj, 132, 2685

\bibitem[\protect\citeauthoryear{{Navarro}, {Frenk} \& {White}}{{Navarro}
  et~al.}{1997}]{NFW97}
{Navarro} J.~F.,  {Frenk} C.~S.,    {White} S.~D.~M.,  1997, \apj, 490, 493

\bibitem[\protect\citeauthoryear{{Press}, {Teukolsky}, {Vetterling} \&
  {Flannery}}{{Press} et~al.}{1992}]{NR}
{Press} W.~H.,  {Teukolsky} S.~A.,  {Vetterling} W.~T.,    {Flannery} B.~P.,
  1992, {Numerical recipes in C. The art of scientific computing}.
Cambridge University Press, 2nd ed.

\bibitem[\protect\citeauthoryear{{Rosati}, {della Ceca}, Burg, Norman \&
  Giacconi}{{Rosati} et~al.}{1995}]{Rosati+95}
{Rosati} P.,  {della Ceca} R.,  Burg R.,  Norman C.,    Giacconi R.,  1995,
  \apj, 445, L11

\bibitem[\protect\citeauthoryear{{Vikhlinin}, {Kravtsov}, {Forman}, {Jones},
  {Markevitch}, {Murray} \& {Van Speybroeck}}{{Vikhlinin}
  et~al.}{2006}]{Vikhlinin+06}
{Vikhlinin} A.,  {Kravtsov} A.,  {Forman} W.,  {Jones} C.,  {Markevitch} M.,
  {Murray} S.~S.,    {Van Speybroeck} L.,  2006, \apj, 640, 691

\bibitem[\protect\citeauthoryear{{Voit}}{{Voit}}{2005}]{Voit05}
{Voit} G.~M.,  2005, Reviews of Modern Physics, 77, 207

\end{thebibliography}

\end{document}